\documentclass[twocolumn,showpacs,preprintnumbers,amsmath,amssymb,aps]{revtex4}
\usepackage{epsf}
\usepackage{graphicx}
\usepackage{dcolumn}
\usepackage{bm}

\begin{document}


\title{Testing of  CP, CPT and causality violation with the
light propagation in vacuum in presence of the uniform
electric and magnetic fields.}

\author{S.L. Cherkas, K.G. Batrakov and D. Matsukevich}
\affiliation{Institute of Nuclear Problems, 220050 Minsk, Belarus}

\date{\today}

\begin {abstract}
We have considered the structure of the fundamental symmetry
violating part of the photon refractive index in  vacuum in the
presence of constant electric and magnetic fields. This part of
the refractive index can, in principle, contain CPT symmetry
breaking terms. Some of the terms violate Lorentz invariance,
whereas  the others violate locality and causality. Estimates of
these effects, using laser experiments are considered.
\end {abstract}


\maketitle
\section {Introduction}
 Recently, the
experiments on searching for the birefringence of a vacuum have
been carried out and planed \cite{camer, pvlas, bmv}. The BMV
project \cite{bmv} was proposed to achieve an accuracy sufficient
for detection of vacuum birefringence, predicted by QED. In
addition, search for exotic non QED interactions is possible in
such experiments. In this article we discuss what kinds of
discrete, i.e., P, T, C symmetries breaking terms can be present
in the photon refractive index in vacuum in constant electric and
magnetic fields.

CP symmetry breaking in $K$- meson \cite{cronin}, and $B$-meson
\cite{babar} decays, as well as time reversal symmetry violation
in $K_0-\bar K_0$ oscillations \cite{cplear} can be currently
described by the standard model with the Cabibbo-Kobayashi-Maskawa
matrix. It would be interesting to find CP violation in the other
systems, different from the $K_0$ or $B_0$. It would be especially
interesting to find some violation of the unconquerable CPT
symmetry. No signals for CPT violation have been observed yet
despite numerous experimental tests.

From the CPT theorem \cite{streater} we know, that CPT invariance
of some field theory
 follows from locality and invariance under Lorenz transforms. Usually  CPT violation is
  considered to be due to  breaking of
Lorentz invariance \cite{kostel}.  However, it is possible, that
locality  is the less fundamental requirement, than Lorentz
invariance, therefore, experiments searching for CPT violation of
both types are of interest.

\section{Electromagnetic wave in vacuum in constant electric and magnetic fields.}
Let us consider the propagation of an electromagnetic wave in
vacuum in  the presence of uniform constant electric and magnetic
fields. Since a photon has no electric charge, such a medium is a
medium with constant refractive index. We will assume that the
field of the electromagnetic wave obeys the Maxwell equation $
\frac {\partial F ^ {\mu\nu} (x)} {\partial x ^\nu} = -4\pi j ^\mu
(x) $, where $j (x) $ is a current of all the particles, which can
interact with the photons. The current arises due to vacuum
polarization by the  electromagnetic wave in presence of the
external fields. Assuming that the wave field is weak and vacuum
in the homogeneous external field remains homogeneous, we can, in
the general case, express the current linearly through the
four-potential of the wave field:
\begin {equation}
j ^ {\mu} (x) = -\int {\mathcal P} ^ {\mu\nu} (x-x') A_\nu (x ')
d^4x',
\end {equation}
where ${\mathcal P} ^ {\mu\nu} (x)$ is some tensor. We do not
consider the case of the strong external electric field, when
vacuum instability \cite{frad} should be taken into account.
 After Fourier transforms
of the four-current $j (x) = \int j (k) e ^ {-ikx} d^4k $ and
four-potential of the electromagnetic field $A (x) = \int A (k) e
^ {-ikx} d^4k $ the Maxwell equation is rewritten as
\begin {equation}
k^2A ^\mu (k) -k ^\mu (kA) = -j ^\mu (k), \label {1}
\end {equation}
where
 \begin {equation} j ^\mu (k) = - {\mathcal P} ^ {\mu\nu} (k) A_\nu (k).
\label {2}
\end {equation}
The four-tensor $ {\mathcal P} ^ {\mu\nu}(k) = \int {\mathcal
P}^{\mu\nu} (x) e ^ {ikx} d^4 x $ should be constructed from the
tensor of the external field ${\mathcal F}^{\mu\nu}$ and a photon
wave vector $k$, since only they are available. By virtue of the
gauge invariance and current conservation the relations $
{\mathcal P} ^ {\mu\nu} k_\nu=k_\mu {\mathcal P} ^ {\mu\nu} =0 $
must be imposed on $ {\mathcal P} ^ {\mu\nu}$.

It is also necessary to emphasize, that all the possible
interactions are supposed to be small, so $k^2$ can be set to zero
on the right-hand side of Maxwell's equation during evaluation of
${\mathcal P} ^ {\mu\nu}$; in addition, only that part of
${\mathcal P} ^ {\mu\nu}$ should be taken into account which does
not become zero when acting on the four-vector polarization
$e_{\mu}(k)$ of a real photon
 (for real photons $k^\mu e_\mu=0$ ).
The structure of the polarization operator, including  off mass
shell terms was considered in Ref. \cite{shabad} for the case when
all the symmetries are conserved and is considered in APPENDIX A
for the case of symmetry violation.

In Eqs. (\ref {1}) and (\ref {2}) the gauge is not fixed yet. We
shall choose the gauge with the null component of the
four-potential being equal to zero: $ \phi=0 $. Then $ \bm E =
-\frac {\partial \bm A} {\partial t} $ and $ \bm E (\bm k, \omega)
=i\omega \bm A $. In a given gauge we obtain from Eqs. (\ref {1})
and (\ref {2}):
 \begin
{equation} \bm k^2 E^i-k^i (\bm k\bm
 E)-\omega^2\left(\delta^{ij}+\frac{{\mathcal
P} ^ {ij}} {\omega^2} \right) E^j=0 \label {main},
\end {equation}
where the three-dimensional tensor $ {\mathcal P} ^ {ij} $ is the
spatial part of the four-tensor $ {\mathcal P}^{\mu\nu} $.
Equation (\ref {main}) shows, that $ \varepsilon ^ {ij} = \delta ^
{ij} + \frac {{\mathcal P} ^ {ij}} {\omega^2} $, plays the role of
the product of the dielectric and magnetic constants the vacuum in
an external field. Further, for short, we shall simply call it the
dielectric constant $\varepsilon^{ij}$ of vacuum in the external
fields.

Let us consider the structure of the four-tensor $ {\mathcal P} ^
{\mu\nu} $ in detail. It can be presented as an expansion in
orders of the external field. Provided the requirements $
{\mathcal P} ^ {\mu\nu} k_\nu=k_\mu {\mathcal P} ^ {\mu\nu} =0 $
\cite{footnote1}, and $e_\mu ^ {*\prime} {\mathcal P} ^ {\mu\nu} e
^\nu\neq 0 $, $k^2=0$ are met, and in the second order in the
external field tensor ${\mathcal F}^{\mu\nu}$ we obtain:
\begin{widetext}
\begin {eqnarray}
{ \mathcal P} ^ {\mu\nu} = a_1 \, {\mathcal F} ^ {\mu\alpha} k_\alpha
{ \mathcal F} ^ {\nu\sigma} k_\sigma +
 a_2\,e^{\mu\lambda\varrho\sigma}{\mathcal
F} _ {\lambda\varrho} k_\sigma e ^ {\nu\phi\delta\alpha} {\mathcal
F} _ {\phi\delta} k_\alpha + i\,b_1 \, e ^ {\mu\nu\alpha\beta}
k_\alpha {\mathcal F} _ {\beta\varrho} {\mathcal F} ^
{\varrho\phi} k_\phi\nonumber
 \\+
c_1 (e ^ {\mu\lambda\varrho\sigma} {\mathcal F} _ {\lambda\varrho}
k_\sigma { \mathcal F} ^ {\nu\delta} k_\delta + e ^
{\nu\lambda\varrho\sigma} {\mathcal F} _ {\lambda\varrho} k_\sigma
{ \mathcal F} ^ {\mu\delta} k_\delta)
 . \label {end1}
\end {eqnarray}
\end{widetext}
 Equation (\ref{end1}) is valid when the external field is slowly varying with
respect to the wavelength of the photon; further terms involving
derivatives of the external field should be included.

  The quantity $e ^ {*\prime} _ \mu {\mathcal P} ^ {\mu\nu}
e_\nu $ is similar to the invariant forward photon scattering
amplitude in the external field. Let us find out its properties
under CPT transformation.  Under  C, P and T  transforms the
tensor of the external field, wave four-vector and
four-polarization of the photon are changed as \cite{lan4}
\begin {eqnarray}
T {\mathcal F} ^ {\mu\nu} \rightarrow - {\mathcal F} _ {\mu\nu}
~~~ T k ^\mu\rightarrow k_\mu ~~~ T e ^\mu\rightarrow e ^
{*\prime} _ \mu ~~~ T e ^ {*\prime} _ \mu\rightarrow e ^
{\mu}\nonumber\\ C {\mathcal F} ^ {\mu\nu} \rightarrow - {\mathcal
F} ^ {\mu\nu} ~~~
C k ^\mu\rightarrow k ^\mu ~~~ C e ^\mu\rightarrow -e ^\mu~~~~~~~~~~~~~~\nonumber \\
P {\mathcal F} ^ {\mu\nu} \rightarrow {\mathcal F} _ {\mu\nu} ~~~
P k ^\mu\rightarrow k_\mu ~~~ P e ^\mu\rightarrow
e_\mu.~~~~~~~~~~~~~~~~~~
\end {eqnarray}
Hence, $ {\mathcal P} ^ {\mu\nu} $ should be symmetric to satisfy
CPT invariance. The term proportional to $b_1$ breaks CPT
invariance with parity breaking only. The term, proportional to
$c_1 $, is CPT invariant, but P-, CP- and T-violating. The terms
proportional to
$a_1=\frac{16}{45}\frac{\alpha^2}{m^4}\approx2.78\times10^{-4}~\mbox{MeV}^{-4}$
and $a_2 =\frac{7}{45}\frac{\alpha^2}{m^4}\approx 1.21\times
10^{-4}~\mbox{MeV}^{-4}$ arise in the framework of  conventional
QED \cite{adler,lan4}. Here $\alpha$ is fine structure constant
and $m$ is the electron mass. From Eq. (\ref {end1}) follows the
explicit form of vacuum dielectric constant \cite{footnote0} in
the stationary uniform electric $ \bm {\mathcal E}$ and magnetic
$\bm { \mathcal B} $ fields:
\begin{widetext}
\begin {eqnarray}
\varepsilon ^ {lj} = \delta ^ {lj} + a_1\biglb( {\mathcal E} ^j
{\mathcal E} ^l + (\bm { \mathcal B} \times \bm n) ^l ( \bm
{\mathcal B} \times\bm n) ^j-{\mathcal E} ^l (\bm {\mathcal B}
\times \bm n) ^j - {\mathcal E} ^j (\bm {\mathcal B} \times \bm n)
^l \bigrb)  +4 a_2\biglb({\mathcal B} ^l {\mathcal B} ^j+ (\bm
{\mathcal E} \times\bm n) ^l {\mathcal B} ^j+ {\mathcal B} ^l (\bm
{\mathcal E} \times \bm n) ^j \nonumber\\+ (\bm {\mathcal E}
\times \bm n) ^l (\bm {\mathcal E} \times \bm n) ^j\bigrb) +i\,b_1
\, e ^ {ljm} \biglb (n^m {\mathcal E} ^2+n^m (\bm {\mathcal E}
\times \bm {\mathcal B} \cdot \bm n) + {\mathcal E} ^m (\bm
{\mathcal E} \bm n) - (\bm {\mathcal B} \times\bm {\mathcal E}) ^m
- (\bm {\mathcal B} \times (\bm {\mathcal B} \times \bm n))
^m\bigrb) \nonumber\\
+2c_1 \, \biglb( {\mathcal B} ^l (\bm {\mathcal B} \times \bm n)
^j + {\mathcal B} ^j (\bm {\mathcal B} \times \bm n) ^l + ( \bm
{\mathcal E} \times\bm n) ^l (\bm {\mathcal B} \times \bm n) ^j +
(\bm {\mathcal B} \times \bm n) ^l (\bm {\mathcal E} \times \bm n)
^j- (\bm {\mathcal E} \times\bm n) ^l {\mathcal E} ^j
\nonumber \\
-{\mathcal E} ^l (\bm {\mathcal E} \times \bm n) ^j - {\mathcal B}
^l {\mathcal E} ^j - {\mathcal B} ^j {\mathcal E} ^l \bigrb) +
id_1 e ^ {ljm} {\mathcal B} ^m +i d_2 e ^ {ljm} {\mathcal E } ^m+i
d_3e ^ {ljm} n^m, \nonumber\\ \label {refr}
\end {eqnarray}
\end{widetext}
where $ \bm n =\frac {\bm k} {\mid \bm k\mid} $ and summation on
the index $m $ is meant. Let us remark, that to an accuracy  up to
the  terms of second order in the external field the refractive
index does not depend on a photon energy (except for the terms
proportional to $d_1$, $d_2$ and $d_3$ about which we can say
nothing). In the Eq. (\ref {refr}) we
 have added "by hands" the terms involving $d_1$, $d_2$ and $d_3$,
which should be absent owing to Lorentz invariance. Such terms as,
for example, the Faraday effect $ \sim e ^ {ljm} {\mathcal B} ^m $
violate both CPT and Lorentz invariance. The same is true for the
term $ \sim e ^ {ljm} {\mathcal E} ^m $, which violates all the
symmetries: P, C, T and Lorentz, although, conserves CP. However,
in the presence of a substance such as gas or plasma they are not
Lorentz violating, because we have additional vector $u^\mu$ of
four-velocity of a substance. The vector $u^\mu$ allows us, for
example, to construct the term ${\mathcal P}^{\mu\nu} \sim
({\mathcal F} ^ {\mu\nu} u ^\eta - {\mathcal F} ^ {\eta\nu} u
^\mu- {\mathcal F} ^ {\mu\eta} u ^\nu) k_\eta $, responsible for
the Faraday effect in a substance.

Therefore,
 experimental detection of the Faraday effect in vacuum
means violation both CPT and Lorentz invariance.

\section {CPT theorem}

 According to the well known CPT theorem \cite{streater}  CPT invariance follows
from  Lorentz invariance and locality, therefore, Lorentz
invariant but CPT violating terms should break locality. Let's
consider it in more detail. A small perturbation of the vacuum in
constant external fields by an electromagnetic wave can be
described in the framework of the Schroedinger formalism (APPENDIX
B). To second order in the current operator $\hat j(x)$ (without
defining its  particular form ) one can obtain that
\begin {eqnarray}
{ \mathcal P} _ {\mu\nu} (x) =4\pi i (< 0\mid \hat j_\mu (x) \hat
j_\nu (0) \mid 0 > ~~~~~\nonumber\\ - < 0\mid \hat j_\nu (0)\hat
j_\mu (x) \mid 0
>) \theta (t), \label {555}
\end {eqnarray}
where $ \theta (t) $ is a step-function. Let's derive
the requirement of CPT invariance again, using a different way. For any
CPT-odd or CPT-even operator $ \hat Z (x) $ one can write:
 $ \Theta ^ {-1} \hat
Z (x) \Theta =\pm\hat Z ^ {+} (-x) $ \cite{streater}, where $
\Theta $  is the operator of CPT reflection, and $ \hat Z ^ {+} $
is a Hermite conjugate operator. Applying this relation to the
product $ \hat j_\mu (x) \hat j_\nu (0) $ and,  taking into
account hermicity
 of the current operator we obtain
\begin {equation}
< 0\mid \hat j_\mu (x) \hat j_\nu (0) \mid 0 > = < 0\mid \hat j_\nu (0) \hat
j_\mu (-x) \mid 0 >, \label {111}
\end {equation}
where invariance of the vacuum under CPT conjugation $\Theta\mid
0>=\mid 0>$ and $<0\mid\Theta^{-1}=<0\mid$
 in the constant uniform field
 is used.
Translational invariance of vacuum in the homogeneous constant field
 $ < 0 |\hat j_\nu (0) \hat j_\mu (-x) |0 > = < 0 |\hat j_\nu (x) \hat j_\mu (0) |0 > $
and Eq. (\ref {111}) lead to the symmetry of the tensor $
{\mathcal P} _ {\mu\nu} (x) = {\mathcal P} _ {\nu\mu} (x) $, as a
condition of CPT  invariance, in agreement with the previous
analysis. In a strong electric field the vacuum is unstable. It
evolves from "empty" state to the state with particle antiparticle
pairs and is no longer T-invariant as well as CPT-invariant. This
leads to the appearance of antisymmetric terms in the polarization
tensor \cite{barash}. The effect is suppressed by multiplier
$e^{-\frac{\pi m^2}{e{\mathcal E}}}$ and negligible for electrons
and laboratory electric field, but what about some unknown light
particles? In the following we will treat vacuum as stable.

First we consider the case, when  the locality condition $< 0
|\hat j ^ {\mu} (x) \hat j ^\nu (0) -\hat j ^\nu (0) \hat j ^\mu
(x) |0 > =0 $ at $x^2 < 0 $ is satisfied.
 This relation implies, that
events at points of four-space, with spacelike separation
 are
not connected in any way. It is a consequence of the limited
velocity of an interaction propagation, or the absence of any
tachyons, which can transfer an interaction.
 Thus, $ {\mathcal P} _ {\mu\nu} (x)
$ is not zero only in the future light cone. Transforms of the
restricted real Lorentz group $L ^\dag _ + $ map the future light
cone $V ^ + $, consisting of points $x^2 > 0$, $x_0 > 0 $, onto
itself \cite {streater}. In other words, the presence of the step
function does not spoil  the Lorentz covariance of ${\mathcal
P}_{\mu\nu}(x)$, as a point of the future with $x_0=t > 0,~x^2>0 $
remains a point of the future with $t ^\prime > 0 $ for any system
of reference.
 This means that the function
\begin {equation}
{ \mathcal P}_{\mu\nu} (k) = \int_{t > 0} {\mathcal P}_{\mu\nu}
(x) e ^{ikx} d^4x. \label {x}
\end {equation}
is covariant under transforms of $L ^\dag _ + $.

The function of Eq. (\ref{x})  can be also defined for the complex
$k = \kappa +i\mathcal {K} $, with $ \mathcal {K}$ belonging to
the future light cone ($ \mathcal {K} \in V ^ + $), since in any
frame of reference $Im \, k_0 > 0 $ and the integral in Eq. (\ref
{x}) converges. Let's define the forward tube $\mathcal F$ as the
set of complex $k =\kappa+i\mathcal {K} $, where $ \mathcal {K} $
belongs to $V ^ +$ \cite {streater}. Then the extended tube $
{\mathcal F}'$ is the set of complex $k$, obtained as a result of
all the complex Lorentz transforms $L _ + (C) $ \cite {streater}
with determinant +1 to the points of $\mathcal F$. Due to
analytical continuation to $ {\mathcal F} ' $ the function $
{\mathcal P}_{\mu\nu}(k)$ becomes covariant under transformations
from the complex Lorentz group $L _ + $. The value of $ {\mathcal
P} _ {\mu\nu}(k)$ at real $k $ is a boundary value of $ {\mathcal
P} _ {\mu\nu} (\kappa) = \lim _ {{\mathcal K} \rightarrow 0,
{\mathcal K} \in V ^ +} {\mathcal P} _ {\mu\nu} (k) $. The
reflections of all four axes is included  into $L _ + $  and,
therefore, $ {\mathcal P}_{\mu\nu} (k) = {\mathcal P}_{\mu\nu}(-k)
$. We can not, however, pass to real $k $ in this equality, as if
in the left-hand side $Im \, k\rightarrow 0 $ belonging $V ^ + $,
in the right-hand side of the equality $Im (-k) \in V ^- $. It is
known that the extended tube contains also real points (Yost
points) \cite{streater}. All Yost points are spacelike. Let's
show, that the relation $ {\mathcal P} _ {\mu\nu} (k) = {\mathcal
P} _ {\nu\mu} (-k) $ is satisfied
 at the Yost points.
 Using the relation $ < 0\mid\hat j_\mu (x) \mid
n> =e ^ {iP_nx} < 0\mid\hat j_\mu (0) \mid n > $
 we rewrite  $ {\mathcal
P} _ {\mu\nu} (x) $ as
\begin{widetext}
\begin {equation}
{ \mathcal P} _ {\mu\nu} (x) =4\pi i\sum_n (< 0\mid\hat j_\mu (0)
\mid n > < n\mid \hat j_\nu (0) \mid 0 > e ^ {-iP_n \, x}
 - <
0\mid \hat j_\nu (0) \mid n > < n\mid\hat  j_\mu (0) \mid 0
> e ^ {iP_n \, x}) \theta (t),
\label {av}
\end {equation}
\end{widetext}

where $P_n\equiv \{\varepsilon_n, \boldsymbol p_n \} $ is the
four-momentum of the particle-antiparticle states in the external
field. Multiplying  Eq.(\ref {av}) by $e ^ {-\epsilon t} $, where
$ \epsilon $ is an infinitesimal number, doesn't spoil convergence
of the integral in Eq.(\ref{x}) and allows us to write $ {\mathcal
P} _ {\mu\nu} (k) $  as
\begin{widetext}
\begin {equation}
{ \mathcal P}_{\mu\nu}(k)=2(2\pi)^4\sum_n\left(\frac{<0\mid \hat
j_\mu (0) \mid n > < n\mid \hat j_\nu (0) \mid
 0>}{\varepsilon_n-\omega-i\epsilon}\delta^{(3)}(\boldsymbol p_n-
\boldsymbol k) + \frac {< 0\mid \hat j_\nu (0) \mid n > < n\mid
\hat j_\mu (0) \mid
 0>}{\varepsilon_n+\omega+i\epsilon}\delta^{(3)}(\boldsymbol p_n +
\boldsymbol k) \right). \label {112}
\end {equation}
\end{widetext}
For the space-like $k $ we can consider  everything in the system of
reference, where $ \omega=0 $, then \begin{widetext}
\begin {eqnarray}
{ \mathcal
 P}_{\mu\nu}(k)=2(2\pi)^4\sum_n\frac{\varepsilon_n}{\varepsilon_n^2 +\epsilon^2}\bigl(<0\mid
\hat j_\mu (0) \mid n > < n\mid\hat  j_\nu (0) \mid 0 > \delta ^
{(3)} (\boldsymbol p_n- \boldsymbol k) \nonumber \\ + < 0\mid \hat
j_\nu (0) \mid n > < n\mid \hat j_\mu (0) \mid 0> \delta ^ {(3)}
(\boldsymbol p_n + \boldsymbol k)  \bigr). \label{13}
\end {eqnarray}
\end{widetext}
From Eq. (\ref{13}) it follows, that the relation $ {\mathcal P} _
{\mu\nu} (k) = {\mathcal P} _ {\nu\mu} (-k) $ is valid.
 By virtue of analytic continuation this
relation  is valid for complex $k $ of the extended tube
${\mathcal F}^\prime$ (though it is violated on passing to the
limit of time-like real $k$ as then in the left-hand side
$Imk\rightarrow 0 $ belongs to the future light cone, while
 in the right-hand side  $Im(-k)$
 approaches zero in the past  light cone). Consequently, we have throughout the extended tube
\begin {equation}
{ \mathcal P} _ {\mu\nu} (k) = {\mathcal P} _ {\nu\mu} (-k) = {\mathcal
P} _ {\nu\mu} (k).
\end {equation}
At  the end and beginning of the equality we can turn to the limit
of real $k $: $Im \, k\rightarrow 0, \, Im \, k \in V ^ + $, and
find that  ${\mathcal P}^{\mu\nu}(k)$ obeys CPT invariance. Thus
we have proved the CPT theorem for our special case, showing that
locality, Lorentz invariance and field-theoretic Schroedinger
equation   lead to the CPT invariance of ${\mathcal
P}^{\mu\nu}(k)$.

Let's assume now, that the local commutativity does not hold for
the operator $\hat j(x)$,  The current operator can be nonlocal
and, for example, may be expressed as $ \hat j_\mu (x) = \int K _
{\mu\nu} (x-x ^\prime) \hat {\mathcal J} ^ \nu (x ^\prime) $,
where the operator $ \hat {\mathcal J} ^ \nu (x) $ is local
(expressed through the fields and their derivatives) and $K _
{\mu\nu} (x-x ^\prime) $ is a function describing nonlocality.
Then, in the general case, the expression of Eq. (\ref {555}) is
distinct from zero at spacelike points. Therefore, due to presence
of the $ \theta$ function Lorentz invariance has been lost. To
maintain the Lorentz invariance we must "remove" the  $
\theta$-function in some way, which will mean violation of
causality. Certainly, we cannot simply remove the $ \theta$
function and are forced to
 abandon the field-theoretic Schroedinger
equation. Modification of the Schroedinger equation to the case of
nonlocal theories is offered in  Ref. \cite{efim}, however, most
likely, it is not unique possibility and we'll not consider it
here.

Thus, experimental detection of the terms of the Faraday effect
type $\sim e ^ {ijm} {\mathcal B} ^m $ and $ \sim e ^ {ijm}
{\mathcal E} ^m $ in vacuum would mean CPT and Lorentz invariance
violation. If we do not detect such terms, but do detect the
CPT-violating term, proportional to $b_1$, it means
 violation of  locality and causality, but Lorentz
 invariance.
Locality and causality may be violated through the presence of
tachyons (particles with the superluminal velocities.) Tachyons
arise in a number of Lorentz invariant theories. Even in the
Rarita-Schwinger theory of a particle of spin 3/2 interacting with
an electromagnetic field tachyonlike solution appears \cite{rar}.
However, no tachyons have been detected experimentally.

 \section {Evolution of light polarization under CP and CPT
violation}

One of the traditional ways to describe light polarization is to
use Stokes parameters $\zeta_1, \zeta_2, \zeta_3$ \cite{gorsh}
which can be measured by experimentalists. We can describe
evolution of the Stokes parameters when  the electromagnetic wave
propagates
 in a medium with a tensor refractive index. Because of the small
difference of the refractive index from unity we can consider
electromagnetic wave to be transverse. Nontransversal terms will
give the next order of smallness in the constants $ a _ 1, a _ 2
\dots $. Thus, the dispersion equation for the wave vector can be
written as:
\begin {equation}
\left (n ^ 2-\hat \epsilon\right) \bm E = 0, \label{29}
\end {equation}
where $ n = k/\omega $, and the wave strength vector $\bm E $ is
perpendicular to $\bm k $ and  has only $ x, y $ components  if
the wave propagates in the $z$-direction. Because of the smallness
$n^2-1$,  Eq. (\ref{29}) can be rewritten as $\left (2n-\hat
\epsilon-1\right)\bm E = 0 $. Putting to zero a determinant of the
equation we can find eigenvectors $ \bm e_l $ belonging to the
eigenvalues $ k_l $.  Expanding the initial strength vector of the
wave $\bm E_0 = \sum_l \alpha_l \bm e_l$ allows one to find the
evolution of the strength vector under the photon propagation
through the volume occupied by the external fields:
\begin {equation}
\bm E (z) = \sum e ^ {ik_lz} \alpha_l e_l = e ^ {i\omega \hat n z}
\bm E _ 0. \label{ez}
\end {equation}
Here we have introduced an operator of the refractive index
according to the formula $ 2 (\hat n -1) = \hat \epsilon -1 $. To
describe partially polarized light the density $ 2\times 2 $
matrix  $ \rho_{ij} = \overline {E_i E^*_j} /\overline {\mid \bm
E\mid ^ 2} $ is used. From Eq. (\ref{ez}) it follows that
\begin {equation}
\frac {d \bm E (z)} {d z} = i\omega \hat n \bm E (z). \label{ez1}
\end {equation}
Eq. (\ref{ez1}) gives the evolution of the density matrix:
\begin {equation}
\frac {d \rho} {d z} = i\omega\left (\hat n\hat\rho-\hat \rho \hat
n^ + -\hat \rho Tr\left \{\hat \rho (\hat n-\hat n ^ +) \right
\}\right). \label {plo}
\end {equation}
Generally, the refractive index operator can be expanded via the
unit  basis vectors $ \bm e _ x $ and $ \bm e _ y $ in the
following way:
\begin {eqnarray}
\hat n-1 = A \bm e _ x\otimes \bm e _ x + B \bm e _ y\otimes \bm e
_ y + C\left (\bm e _ x\otimes \bm e _ y + \bm e _ y\otimes \bm e_
x\right) \nonumber\\ + iD\left (\bm e _ x\otimes \bm e _ y-\bm e _
y\otimes \bm e _ x\right),~~~
\end {eqnarray}
or in the  matrix form
\begin {eqnarray}
\hat n -1 = A\left (\begin {array} {cc} 1 & 0 \\ 0 & 0
\end {array} \right) + B\left (\begin {array} {cc} 0 & 0 \\ 0 & 1
\end {array} \right) + C\left (\begin {array} {cc} 0 & 1 \\ 1 & 0
\end {array} \right)
\nonumber\\
 +D\left (\begin {array} {cc} 0 & i \\ -i & 0
\end {array} \right),
 \label {refstr}
\end {eqnarray}
where quantities $A, B, C, D$ should be expressed in terms of
$a_1, a_2 \dots$ for a concrete external field configuration.

 The
density matrix can be parameterized by the Stokes parameters
\cite{gorsh} :
\begin {equation}
\hat \rho = \frac {1} {2} \left (
\begin {array} {cc}
1 + \zeta _ 3 & \zeta _ 1-i\zeta _ 2 \\
\zeta _ 1 + i\zeta _ 2 & 1-\zeta _ 3
\end {array}
\right).
\end {equation}
Distinguishing the real and imaginary parts in the coefficients
 $
A = A ^ \prime + iA ^ {\prime\prime} \dots $ we find from the
equation (\ref {plo}) that
\begin{widetext}
\begin {eqnarray} \frac {1} {\omega} \frac
{d\zeta _ 1} {dz} = (A '-B ') \zeta _ 2 + ( A ^ {\prime\prime} -B
^ {\prime\prime}) \zeta _ 1\zeta _ 3-2C ^ {\prime\prime} (1-\zeta
_ 1 ^ 2) + 2D '\zeta _ 3-2D ^ {\prime\prime} \zeta _ 1\zeta _ 2
\nonumber \\
\frac {1} {\omega} \frac {d\zeta _ 2} {dz} = (B '-A ') \zeta _ 1 +
( A ^ {\prime\prime} -B ^ {\prime\prime}) \zeta _ 2\zeta _ 3 + 2C
'\zeta _ 3 + 2C ^ {\prime\prime} \zeta _ 1\zeta _ 2 + 2D ^
{\prime\prime} (1-\zeta _ 2 ^ 2)
\nonumber \\
\frac {1} {\omega} \frac {d\zeta _ 3} {dz} = - (A ^ {\prime\prime}
-B ^ {\prime\prime}) (1-\zeta _ 3 ^ 2) -2C '\zeta _ 2 + 2C ^
{\prime\prime} \zeta _ 1\zeta _ 3-2D '\zeta _ 1 + 2D ^
{\prime\prime} \zeta _ 2\zeta _ 3.
\end {eqnarray}
\end{widetext}

\begin{figure}[h]
\hspace {-2. cm} \epsfxsize =5. cm \epsfbox[50 370 410
650]{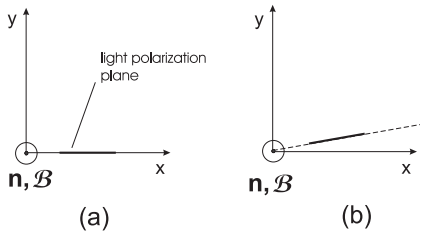} \caption{(a) Linearly polarized light with
$\zeta_3=1$, $\zeta_1=\zeta_2=0$, (b) the light with
$\zeta_3\approx 1$, $\zeta_1\ne 0$, $\zeta_2=0$.}
 \vspace{-0.5cm}
\label{stoks0}
\end{figure}

The Faraday effect can be measured, if we choose the magnetic
field to be parallel to the wave vector of the photon as shown in
Figs. \ref{stoks0}(a), (b). Then in Eq. (\ref{refstr}) the only
term proportional to $D=\frac{1}{2}(d_1\mathcal{B}+d_3)$ remains.
As light passage through the volume occupied by the magnetic
field, the light with the only initially distinct from zero Stokes
parameter $ \zeta_3 $ \cite{gorsh} will gain polarization
corresponding to the parameter $ \zeta_1 $ and, in contrast, light
with the only initially non-zero parameter $ \zeta_1 $ gains
polarization corresponding to $ \zeta_3$. Thus the light
polarization rotates as it is shown in Fig. \ref{stoks0}.

Let us recall that
 $\zeta_3=1$ and $ \zeta_3=-1$  corresponds to the light polarized
along $x$ and $y$ axis respectively.  The parameter $\zeta_1=\pm
1$ describes polarization at $45^{0}$ to the $y$ axis.  The light
ellipticity $\Psi$ is expressed through $\zeta_2=2\Psi$ for fully
polarized light. Partially polarized light can be expanded as sum
of natural light and elliptically polarized light. In this case
$\zeta_2=2\Psi \,P$, where $P$ is light polarization and $\Psi$ is
the ellipticity of the polarized part.

In the BMV project  \cite{bmv} it is planed to achieve an accuracy
sufficient for the measurements of the vacuum birefringence
predicted by QED, i.e., $ \Delta n\sim 10 ^{-21} $ at ${\mathcal
B}\sim 25~~ \mbox{T}$. Earlier,  $ \Delta n\sim 1.3\times 10
^{-20}$ was measured in the BNL experiment \cite{camer} and $
\Delta n\sim 6.7\times 10 ^{-20}$ by PVLAS \cite{pvlas}. Thus,
from measurements of $\Delta n$ corresponding to Faraday rotation
at the level $ \Delta n_{CPTL}=D\sim 10 ^{-21}$ in a magnetic
field of 25 T (we'll use the system of units $e^2/4\pi=\alpha, 1\,
\mbox{T}=195\, \mbox{eV}^2, 1
\,\mbox{V}/\mbox{m}=6.5\times10^{-7}\, \mbox{eV}^2$) one will be
able to obtain the restriction $d_1=\frac{2D}{\mathcal B}\sim
4\times 10^{-13}~\mbox{MeV}^{-2}$.

The presence  of a residual pressure in the resonator imposes
restriction on the measurement of $ \Delta n $ of the vacuum.
Assuming the residual pressure in the equipment to be $ 10 ^ {-11}
$ Torr, we find that $ \Delta n $ of the Faraday effect for helium
at this pressure is $ \Delta n\sim 10 ^ {-22} $. Thus,
CPT-violating Faraday effect can be measured with this accuracy.

\begin{figure}[h]
\hspace {-2. cm} \epsfxsize =5. cm \epsfbox[50 370 410
650]{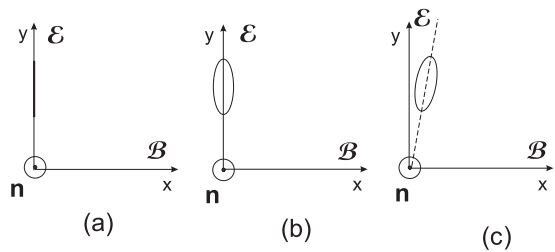} \caption{(a) Linearly polarized light with
$\zeta_3=-1$, $\zeta_1=\zeta_2=0$, (b) the light with
$\zeta_3\approx -1$, $\zeta_2\ne 0$ and $\zeta_1=0$, (c) the light
with $\zeta_3\approx -1 $, $\zeta_2\ne 0$ and $\zeta_1\ne 0$
(ellipse of polarization is slightly rotated). }
 \vspace{-0.5cm}
\label{stoks}
\end{figure}

\begin{figure}[h] \hspace{-2.5 cm} \vspace{0.0cm} \epsfxsize =4. cm
\epsfbox[0 120 310 650]{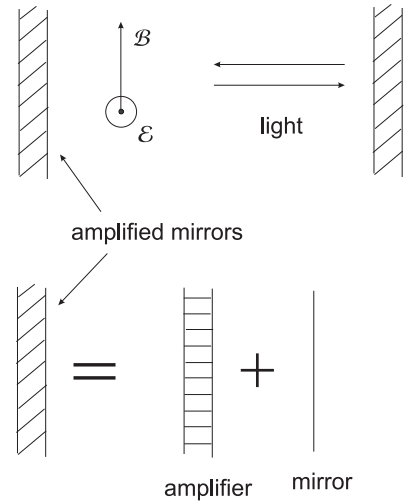} \vspace{1.cm} \caption{Scheme of
a photon trap.} \label{ampl}
\end{figure}

For measurement of the terms, proportional to $b_1 $  we may
choose a magnetic field perpendicular to the photon wave vector.
The electric field should be chosen perpendicular to both the
photon wave vector and the magnetic field strength vector. Thus
the photon wave vector, the
 direction of the magnetic field, and the direction
of the electric field form triplet of mutually orthogonal vectors
as shown in Fig. \ref{stoks} and \ref{ampl}. The refractive index
contains terms which are of odd or even order in the vector $\bm
n$. For the laser experiment only the terms of  even order in the
wave vector are of interest, because these effects accumulate
under the passage of a photon back and forth between the resonator
mirrors \cite{footnote2}. Considering only these terms we find
that all the coefficients $ A=\frac{a_1}{2}({\mathcal
E}^2+{\mathcal B}^2)$, $B=2a_2({\mathcal E}^2+{\mathcal B}^2)$,
$C=-c_1{\mathcal E}{\mathcal B}$ and $D=-b_1{\mathcal E}{\mathcal
B}+d_3/2$ are different from zero. First, assume that all the
coefficients are approximately of the same order of magnitude.
Then the light with the initial polarization $ \zeta_3$ receives
polarization $ \zeta_1 $. The polarization $ \zeta_1 $ can also
arise due to imaginary part of the coefficient $C$, however, this
contribution does not depend on the sign of the initial
polarization $ \zeta_3 $ and can be separated by changing the sign
of $ \zeta_3 $ during the experiment. Taking the electric field
strength $\mathcal{ E}\sim 10^{6}$ V/m we obtain a restriction on
CPT and the causality violation constant $b_1=\frac{\Delta
n^\prime_{CPT}}{\mathcal{E}\mathcal{B}}=\frac{D}{\mathcal{E}\mathcal{B}}\sim
0.31 \,\mbox{MeV}^{-4} $ if $\Delta n^\prime_{CPT}$ is measured
with accuracy $10^{-21}$. To measure CP-violating constant $c_1$
we have to search for the ellipticity parameter $\zeta_2$ when the
light was initially linearly  polarized with the $\zeta_3=1$.

In the case, when $ \mid A-B\mid \gg \mid D \mid, \mid C\mid$ (but
 $\mid (A'-B') \omega z\mid\ll 1 $)
 "mixing" of the polarizations $ \zeta_1 $ and $ \zeta_2 $ occurs. Still,
the light initially polarized with $ \zeta_3 $ will gain
polarization $ \zeta_1 $ and $ \zeta_2 $ only in the case, when $C
$ or $D$ differs from zero. But we will not know $C $ or $D $.
Fortunately, we have a possibility to avoid this difficulty.
 The sign of the Cotton-Mouton effect  for nitrogen is
opposite to the sign of the vacuum Cotton-Mouton effect;
therefore, using nitrogen at a residual pressure about $ 10 ^ {-7}
$ Torr we can compensate for the difference $A'-B'$ and
distinguish $D$ from $C$.

 Apparently, the possibility exists to measure much smaller $
\Delta n $. Baryshevsky offers an interesting idea of using laser
amplifiers \cite{bar}, which do not change polarization properties
of  light, but at the same time, will stop a photon beam damping.
Ideally, the amplifier should be combined with a mirror, as shown
in Fig. \ref{ampl}, to obtain  "amplified" mirror with the
reflectivity 1 or more than 1. A light can be localized in such a
trap for several hours. Assuming, for example, that we can measure
an angle of polarization rotation $ \Delta \theta =\Delta n \omega
z \sim 10^{-10} $ and the lifetime of a photon in the trap is 1
hour, we'll find the minimum measured $ \Delta n\sim 10^{-27} $,
for the $\omega=2.4 ~eV$. However a number of technical problems
can arise in this scheme. For instance, we need the amplifier
remaining isotropic after multiple passage of the polarized light
through it.

Finally it may be possible  to obtain restrictions on this CPT
violating term  examining the polarization of light, from
 the distant galaxies. It is necessary to separate the vacuum
effects from the Faraday rotation in magnetic field and substance
of galaxies. Earlier, such an analysis yields the restriction
$\Delta n_{CS}=d_3 \sim 10 ^ {-33}$ ($\omega=2.4 ~eV$)
 \cite{jackiw} for the term $\varepsilon^{ij} \sim
e ^ {ijm} n^m $ (Chern-Simon term).
\begin {figure}[h]
\includegraphics {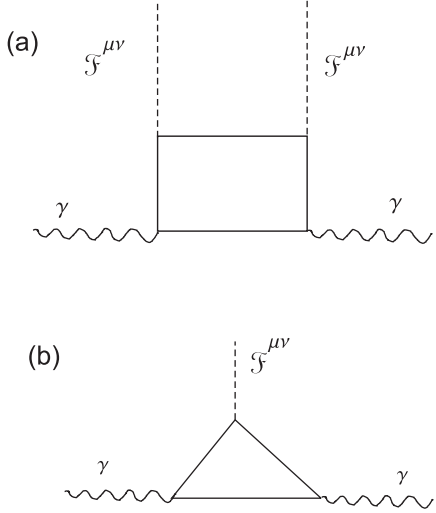}
\caption{(a) QED graph of vacuum polarization contribution to the
photon refractive index in a static field. (b) Analogous graph of
C-odd vacuum polarization.} \label{cpt}
\end {figure}

\section{Comparison of the laser experiment tests with some other
known tests } 
\begin {figure}[t]
\includegraphics{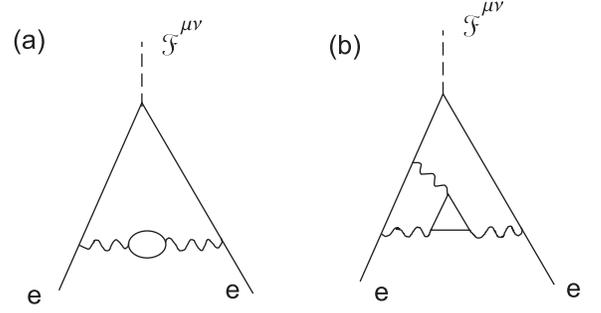}
\caption{(a) Graph of vacuum polarization contribution to the
electron or positron g-factor. (b) Analogous graph of C-odd vacuum
polarization.} \vspace{0. cm} \label{gf}
\end {figure}

Let us estimate CPT violation of the Faraday type $
\sim e ^ {ijm} {\mathcal B} ^m$. Certainly, we can not be sure of
the  applicability of the Feynman diagram technique in the case of
CPT invariance violation. But it can still be suitable for
heuristic estimates. In the framework of  QED, the refractive
index, proportional to $a_1, a_2 $, is evaluated using the square
diagram shown in Fig. \ref{cpt}. Each vertex with the external
electromagnetic field corresponds to  the factor
$\frac{e}{\sqrt{4\pi}} \frac { {\mathcal B}} {m^2} $ or
$\frac{e}{\sqrt{4\pi}} \frac { {\mathcal E}} {m^2} $ in $\Delta
n$, where $m$ is the electron mass.   Each of the remaining
vertices corresponds to the factor $e/\sqrt{4\pi}$. We can, in the
same way, estimate CPT and Lorentz violating term $ \sim e^{ijm}
{\mathcal B} ^m$, considering the triangle diagram shown in Fig.
\ref{cpt}. The triangle diagram can not appear in  standard QED,
as the diagram is not invariant under C conjugation. Let us assume
the most remarkable possibility, that the violation of C, CPT and
Lorentz invariance is induced by some unknown particles
interacting with the photons with C violation of the order of
unity.
 By analogy with the calculation of the standard square diagram
we assume that the vertex  with the external field corresponds to
the factor $\frac{g}{\sqrt{4\pi}} \frac { {\mathcal B}} {\mu ^ 2}$
in $\Delta n$ and the others vertexes correspond to the factor
$g/\sqrt{4\pi}$ in $\Delta n$, where $g$ is the coupling of the
particle with photons and $\mu$ is the particle mass. As a result,
we have
\begin {equation}
\Delta n _ {CPTL} \approx \frac{{g} ^ 2}{4\pi} \, \frac {{ g}
{\mathcal B}} {\sqrt{4\pi}\mu ^ 2}.
\end {equation}
In the BMV project it is planed to reach an accuracy sufficient
for a measurement of $ \Delta n $ predicted by QED,  i.e., $\Delta
n\sim 10 ^{-21}$ (strength of the magnetic field is 25 T ) A
measurement of $\Delta n _ {CPTL}$ with  such an accuracy gives a
restriction on the coupling $g$. It is interesting to compare this
restriction with what follows from the CPT test, based on a
comparison of the $\mbox{ g} $ factors of an electron and
positron: $ \alpha _ {CPT} = \frac {\mbox{g} _ {e ^ +} -\mbox{g} _
{e ^ -}} {\mbox{g} _ {avr}} < 10 ^ {-12}$ \cite{part}. Under the
assumption of the same mechanism of C parity violation, $ \alpha _
{CPT}$ arises from the diagram shown in Fig. \ref{gf}(b).
For the sake of simplicity, we again make very heuristic estimates
of the diagram shown in Fig. \ref{gf}(b). First, we remark that
the relative contribution of the diagram shown in Fig. \ref{gf}(a)
(usual vacuum polarization) to the g-factor of the electron is $
\sim \frac{e ^ 4}{(4\pi)^2} \frac {m ^ 2} {\mu ^ 2} $ if a virtual
particle mass  $ \mu\gg m$, and is $ \sim \frac{e ^ 4}{(4\pi)^2}
\ln {\frac {m} {\mu}} $ when $ \mu\ll m$ \cite{lan4}. This fact is
a reflection of a more general rule. The contribution of the
virtual particle loop connected by the photon lines to the
electrons is proportional to some degrees of $ \frac {m} {\mu} $
when $ \mu\gg m$, and to some degrees of $ \ln{\frac {m} {\mu}} $
when $ \mu \ll m $. Thus, the contribution of the diagram shown in
Fig. \ref{gf}(b) can be estimated as
\begin {equation}
\alpha _ {CPT} \approx \frac{e ^ 3 {g} ^ 3}{(4\pi)^3} F\left
(\frac {m ^ 3} {\mu ^ 3} \right),
\end {equation}
where the Spens function  $ F(x)$ \cite{lan4} has the asymptotic $
F(x) \approx x $ when $ x\ll 1 $ and $ F(x) \approx \frac {\pi ^
2} {6} + \frac {1} {2} \ln ^ 2(x)$ when $ x\gg 1 $.
\begin{figure}[h]
\hspace{-2.5cm} \vspace{0.0cm} \epsfxsize =6. cm \epsfbox[0 50 410
650]{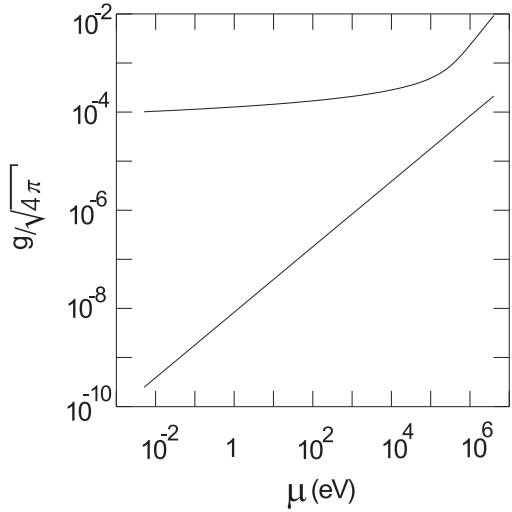} \vspace{-1.5 cm}   \caption{Restrictions on the
dimensionless coupling $g$ of some particles interacting with
photons, with the relative C, CP, CPT and Lorentz invariance
violation of order unity. The straight curve relates the Faraday
effect in vacuum and corresponds to the inequality $\Delta
n_{CPTL} <10^{-21}$.  The bent curve arises from the difference of
electron and positron g-factors and corresponds to the inequality
$\frac {\mbox{g} _ {e ^ +} -\mbox{g} _ {e ^ -}} {\mbox{g} _ {avr}}
< 10 ^ {-12}$. } \label{g1}
\end{figure}
Fig. \ref{g1} shows restrictions on the coupling $g$ of C-, CP -,
CPT- and Lorentz-violating interaction following from the
inequalities $ \Delta n _ {CPTL} < 10 ^ {-21} $ and $ \alpha _
{CPT} < 10 ^ {-12} $. As we can see, measurement of the Faraday
effect in vacuum gives much more stringent restrictions on $g$  in
the above model of CPT violation than the traditional comparison
of the electron and positron g-factors. Certainly, it happens
because we have chosen the model with CPT violation in the photon
sector; therefore, the experiments  dealing directly with photons
have an advantage.

Let us now consider the terms proportional to $ b _ 1 $, which
breaks P , CP , CPT and causality, but at the same time, are
Lorentz invariant and  do not break C-parity. To estimate of the
appropriate $\Delta n _ {CPT}^ \prime$ we should consider the
square diagram of Fig. \ref{cpt}(a). In the same way we find:
\begin {equation}
\Delta n _ {CPT} ^ \prime\sim \frac{{g} ^{ \prime 2}}{4\pi} \,
\left (\frac {{ g} ^ \prime {\mathcal B}} {\sqrt{4\pi}\mu ^2}
\right) \left (\frac {{g} ^ \prime {\mathcal E}}
{\sqrt{4\pi}\mu^2} \right).
\end {equation}
The restriction on $ {g}'$, obtained from a measurement of $
\Delta n_ {CPT}^ \prime $ with the accuracy $ 10 ^ {-21} $, can be
compared, for example, with the restriction on CP  violation in
para-positronium decay in two photons. Positronium $ ^ 1S _ 0 $
state has negative spatial parity \cite{lan4}; therefore, the
probability of decay into two polarized photons should be
proportional to $ (\bm e _ 1\times \bm e _ 2\cdot \bm k) $
\cite{perk}, where $ \bm e _ 1 $ and $ \bm e _ 2 $ are the photon
polarizations, $\bm k$ is the momentum of one of the photons
(another photon has opposite momentum). The presence of the P-even
$ (\bm e _ 1\cdot\bm e _ 2) $ correlation is a signal of P and CP
violation (C parity conserves in para-positronium two photon
decay). For this process we can say nothing about  T invariance,
because we do not compare it with the reverse process of the
$\gamma+\gamma\rightarrow Ps$.
\begin {figure} [h]
\vspace{0.5 cm}
\includegraphics {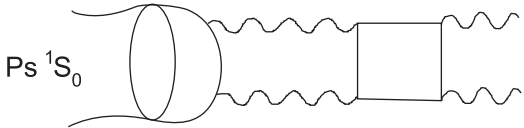}
\caption{Graph of P-violating photon interaction in final state in
the two photon para-positronium decay. } \label{pos}
\end {figure}
The branching ratio $ \alpha _ {CP} ^ \prime $ of the decay with
the $ (\bm e _ 1\cdot\bm e_ 2) $ can be estimated from the diagram
shown in Fig. \ref{pos} and is given by
\begin {equation}
\alpha _ {CP} ^ \prime\sim \frac{g ^ {\prime 4}}{(4\pi)^2} F\left
(\frac {m ^ 4} {\mu^ 4} \right).
\end {equation}
\begin{figure}[h]
 \hspace{-2.cm} \vspace{0. cm}
\epsfxsize =6. cm \epsfbox[0 200 410 650]{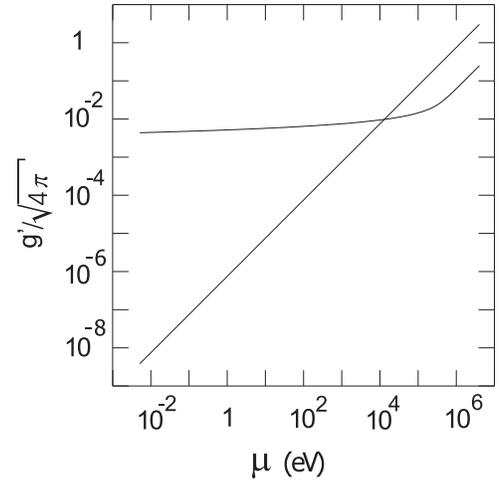} \vspace{0. cm}
 \caption{Restrictions on the
dimensionless coupling $g^\prime$ of the particles interacting
with photons with  violation of parity, CP, CPT and causality of
an order unity, but conserving Lorentz invariance. The
restrictions follow from the inequality $\Delta n _ {CPT}^\prime <
10 ^ {-21}$ (straight curve) and the inequality $\alpha _
{CP}^\prime < 10 ^ {-6}$ (bent curve).
 }
\label{g22}
\end{figure}

\begin {figure} [!]
\includegraphics{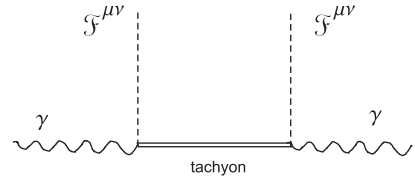}
\vspace{0.9 cm} \caption{CPT violation due to exchange of
causality violating particle (tachyon) of  axion type.}
\label{cpt_ax}
\end {figure}

The restrictions on $ {g} ^ \prime$, following from the inequality
$\Delta n _ {CPT}^\prime < 10 ^ {-21}$ and $\alpha _ {CP}^\prime <
10 ^ {-6}$ are shown in Fig.\ref{g22}.
 We have taken the electric field strength $10 ^ 6$ V/m,
so that the dimensionless parameter $ \left (\frac {e {\mathcal
E}} {m ^ 2} \right) \sim 10 ^ {-12} $. The vertexes, we have
considered, are  of  the type one photon --- two particles,
however, it is  possible to consider a vertex  of the type two
photons
--- one particle.  In this case for evaluation of $ \Delta n _
{CPT} ^ {\prime\prime}$ we should to consider the diagram shown in
Fig. \ref{cpt_ax}. From the reasons of dimensionality  we obtain:

\begin {equation}
\Delta n _ {CPT}^{\prime\prime} \sim\left (\frac
{{g}^{\prime\prime} {\mathcal B}} {\sqrt{4\pi}\mu ^ 2} \right)
\left (\frac {{g} ^ {\prime\prime} {\mathcal E}} {\sqrt{4\pi}\mu^
2} \right).
\end {equation}

\begin {figure} [h]
\includegraphics{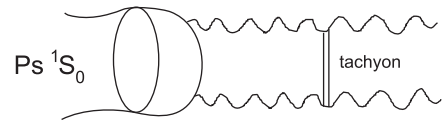}
\caption{Tachyon induced CP violation in the two-photon decay of
para-positronium. } \label{pos_ax}
\end {figure}
CP violation in positronium decay can be estimated from the
diagram shown in Fig. \ref{pos_ax}  as
\begin {equation}
\alpha _ {CP}^{\prime\prime}\sim \frac{{g}^{\prime\prime 2}}{4\pi}
F\left (\frac {m^2} {\mu^2} \right).
\end {equation}
\begin{figure}[h]
\hspace {-2.5 cm} \epsfxsize =6. cm \epsfbox[0 20 410 680]{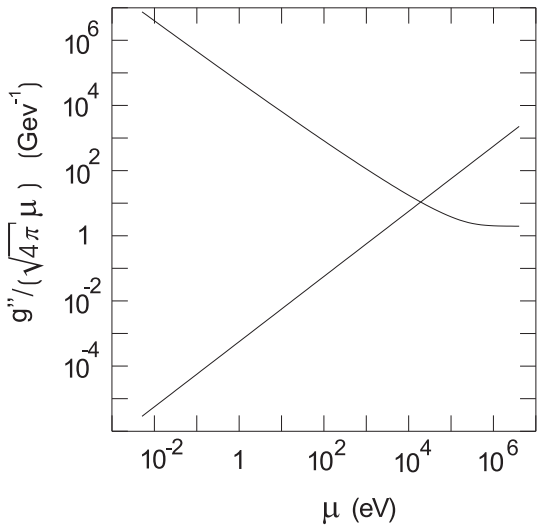}
\vspace{-0.5 cm} \caption{Restrictions on the tachyon coupling
arising from the inequality $\Delta
n^{\prime\prime}_{CPT}<10^{-21}$ (straight curve) and from the
inequality $\alpha_{CP}^{\prime\prime}<10^{-6}$ (bent curve). }
\label{g3}
\end{figure}

Unfortunately, due to the weakness of the electric field possible
in a laser experiment, compared to the magnetic one, restriction
on a such CPT and causality breaking tachyon coupling are  $ \sqrt
{\frac {\mathcal E} {\mathcal B}} $ times weaker than the
restriction on the usual axion coupling for which $ \Delta
n\sim\left (\frac {{ g} _ a {\mathcal B}} {\sqrt{4\pi}\mu^ 2}
\right) ^ 2 $. For the strength of the electric and magnetic
fields, used in our work,  this gives about 100 times difference.
For the usual axion the very rigid restriction
$\frac{g_a}{\sqrt{4\pi}\mu_a}\sim
10^{-9}-10^{-10}~~\mbox{Gev}^{-1}$ follow from astrophysics
\cite{raf}. However, laser experiments can be considered
irrespective of the models as  independent tests of CPT
invariance.

\section{Conclusion}
To summarize, laser experiments on searching CPT, Lorentz
invariance and causality violation for photons in vacuum, in the
presence of constant uniform magnetic and electrical fields, are
competitive with  tests using positron and electron g-factors
comparison and searching of CP violation in positronium decay,
provided, that CPT is broken in the photon sector. It is
essential, that in the case of unbroken Lorentz invariance we have
the possibility of testing causality and locality.

\section*{ACKNOWLEDGMENT}
The authors are grateful to Prof. Vladimir Baryshevsky for
valuable discussions.

\appendix
\section*{APPENDIX A}
In this appendix we consider the structure of the tensor
${\mathcal P}^{\mu\nu}$ in the external field, including of  mass
shell terms. Despite the above classical consideration the
expansion of the tensor ${\mathcal P}^{\mu\nu}$ in the orders of
${\mathcal F}^{\mu\nu}$ and $k^\mu$ is  not only valid for  soft
photons.
  In fact we can deduce it by
considering one-photon retarded Green's function in the external
stationary uniform electromagnetic field ${\mathcal
D}_{\mu\nu}(x-x^\prime)= i<0\mid[\hat A_{\mu}(x), \hat
A_\nu(x^\prime)]\mid 0>\theta(t-t^\prime)$. The photon propagation
modes can be described by the poles of the Fourier transform
${\mathcal D}_{\mu\nu}(k)=\int {\mathcal
D}_{\mu\nu}(x)e^{ikx}d^4x$ of the Green's function.
  The dispersion relation for the propagation modes reads
\begin{equation}
det\mid {\mathcal D^{-1}_{\mu\nu}(k)}\mid=0.
\end{equation}
The photon Green's function is expressed through the Green's
function of the free photon $D_{\mu\nu}(k)$ and the polarization
operator as
\[{\mathcal
D}_{\mu\nu}(k)=D_{\mu\nu}(k)+D_{\mu\alpha}(k){\mathcal
P}^{\alpha\beta}(k){\mathcal D}_{\beta\nu}(k).
\]
 Thus ${\mathcal
D}_{\mu\nu}^{-1}=D_{\mu\nu}^{-1}-{\mathcal P}^{\mu\nu}.$ Taking
$D_{\mu\nu}^{-1}(k)=\left(k^2g_{\mu\nu}-k_\nu k_\mu\right)$ we
come to a dispersion relation congruous to Eqs.(\ref{1}) and
(\ref{2}).

Tensor  4x4 ${\mathcal P}^{\mu\nu}$ contains 16 independent
components. Thus it can be expanded over 16 independent tensors.
Ten of them are symmetric and 6 are antisymmetric. The gauge
invariance condition $ {\mathcal P}^{\mu\nu}k_\nu, ~~\mu
\{0,1,2,3\}$ reduces the number of symmetric terms to 6. It also
reduces the number of antisymmetric terms to 3, because for any
antisymmetric tensor $k_\mu{\mathcal P}^{\mu\nu}k_\nu=0$  is
automatically valid and only three of four gauge conditions are
independent. Independent tensors should be expressed through the
tensor of an external field ${\mathcal F}^{\mu\nu}$ and the photon
wave vector $k$. The expansion can be written as
\begin{widetext}
\begin{eqnarray}
{ \mathcal P}^{\mu\nu} =a_0(k^2g^{\mu\nu}-k^\mu k^\nu) +
 a_1  {\mathcal F} ^ {\mu\alpha} k_\alpha
{ \mathcal F}^{\nu\sigma} k_\sigma +
 4a_2 \tilde{\mathcal F} ^ {\mu\alpha} k_\alpha
\tilde{ \mathcal F}^{\nu\sigma} k_\sigma + 2c_1(\tilde {\mathcal
F} ^ {\mu\alpha} k_\alpha { \mathcal F} ^ {\nu\sigma}
k_\sigma+{\mathcal F}^{\mu\alpha} k_\alpha \tilde{ \mathcal F} ^
{\nu\sigma} k_\sigma)\nonumber \\+c_2[(k^2{\mathcal
F}^{\mu\alpha}{\mathcal F}_{\alpha\lambda}k^\lambda -k^\mu
k^\delta {\mathcal F}_{\delta\beta}{\mathcal
F}^{\beta\sigma}k_\sigma){\mathcal F}^{\nu\rho}k_\rho+
(k^2{\mathcal F}^{\nu\alpha}{\mathcal F}_{\alpha\lambda}k^\lambda
-k^\mu k^\delta {\mathcal F}_{\delta\beta}{\mathcal
F}^{\beta\sigma}k_\sigma){\mathcal F}^{\mu\rho}k_\rho]
\nonumber\\
+c_3[(k^2{\mathcal F}^{\mu\alpha}{\mathcal
F}_{\alpha\lambda}k^\lambda -k^\mu k^\delta {\mathcal
F}_{\delta\beta}{\mathcal F}^{\beta\sigma}k_\sigma)\tilde{\mathcal
F}^{\nu\rho}k_\rho+(k^2{\mathcal F}^{\nu\alpha}{\mathcal
F}_{\alpha\lambda}k^\lambda -k^\mu k^\delta {\mathcal
F}_{\delta\beta}{\mathcal F}^{\beta\sigma}k_\sigma)\tilde{\mathcal
F}^{\mu\rho}k_\rho]
\nonumber\\
+b_1\,  e^{\mu\nu\alpha\beta} k_\alpha {\mathcal F} _ {\beta\eta}
{\mathcal F}^{\eta\phi} k_\phi+ b_2\,
e^{\mu\nu\lambda\alpha}{\mathcal
F}_{\lambda\sigma}k^{\sigma}k_{\alpha}+ b_3\, (k^\mu{\mathcal
F}^{\nu\lambda}k_\lambda-k^\nu{\mathcal
F}^{\mu\lambda}k_\lambda+k^2{\mathcal F}^{\mu\nu}), \label{end2}
\end {eqnarray}
\end{widetext}
where $\tilde{\mathcal
F}^{\mu\nu}=\frac{1}{2}e^{\mu\nu\eta\delta}{\mathcal
F}_{\eta\delta}$. Coefficients $a_0, a_1, \dots$ are functions of
four independent scalars $k^2$, $k_\mu {\mathcal
F}^{\mu\nu}{\mathcal F}_{\nu\lambda}k^\lambda$, $\Im={\mathcal
F}_{\mu\nu}{\mathcal F}^{\mu\nu}$, ${\mathcal G}={\mathcal
F}_{\mu\nu}\tilde{\mathcal F}^{\mu\nu}$ \cite{shabad}. The terms
involving $a_0, c_2, c_3, b_2, b_3$ do not lead to  observable
effects at first order in the constants, because evaluation of
these  terms on the photon mass shell gives zero. For instance,
the quantity $e^{*\mu}(k^2g_{\mu\nu}-k_\mu k_\nu)e^{\nu}$ equals
to zero because the free photon satisfies $k^2=0$ and $ek=0$. The
symmetry properties of the all the terms are given in the table.
Let us remark that the scalar $\mathcal G$ is P- and T- violating
so if the coefficients $a_0, a_1 \dots$ contain odd orders of
$\mathcal G$ its symmetry properties change. Conventional QED
allows the terms proportional to $a_0, a_1, a_2$ and also the term
involving $c_1$ which appears only with the odd degree of
${\mathcal G}$. In a pure magnetic field ${\mathcal G}=0$ and the
aforementioned term disappears.
\begin{table}
\caption{\label{tab:table3}Symmetry properties of the terms of
tensor ${\mathcal P}^{\mu\nu}$ allowed by Lorentz and gauge
invariance.}
\begin{ruledtabular}
\begin{tabular}{ccccc}
 &\multicolumn{2}{c}{Symmetry$~~~~~~~$}&Observability\\
 Term&Base&Modified by ${\mathcal G}^{2n+1}$&with real\\
&C P T&C P T&photons\\
\hline
 $a_0$&$+$ $+$ $+$&$+$ $-$ $-$ &invisible\\
 $a_1$&$+$ $+$ $+$&$+$ $-$ $-$ &visible\\
 $a_2$&$+$ $+$ $+$&$+$ $-$ $-$ & $<>$\\
 $c_1$&$+$ $-$ $-$&$+$ $+$ $+$ & $<>$\\
 $c_2$&$-$ $+$ $-$&$-$ $-$ $+$ &invisible\\
 $c_3$&$-$ $-$ $+$&$-$ $+$ $-$ &$<>$\\
 $b_1$&$+$ $-$ $+$&$+$ $+$ $-$ &visible \\
 $b_2$&$-$ $+$ $+$&$-$ $-$ $-$ &invisible\\
 $b_3$&$-$ $-$ $-$&$-$ $+$ $+$ &$<>$\\
\end{tabular}
\end{ruledtabular}
\end{table}

\appendix
\section*{APPENDIX B}

Here we deduce Eq.(\ref{555}) from the field-theoretic
Schroedinger equation. The classical 4-current $ j_ \mu (\bm r, t)
$ corresponds to some Schroedinger operator $\hat j _ \mu (\bm
r)$, so that perturbation of the vacuum in a constant fields by
the electromagnetic wave can be described by the interaction
Hamiltonian $ \hat V (t) = \int \hat j ^ \mu (\bm r) A _ \mu (x) d
^ 3r $. Let's recall that $ A(x)$ represents the 4-potential of
the wave. We'll assume that the  wave rise adiabatically from zero
value at infinity. A perturbed state of the system is described by
the Schroedinger equation.
\begin {equation}
\frac{d}{dt} \mid t> = (\hat H _ 0 + \hat V) \mid t>. \label {sh1}
\end {equation}
 Vacuum states in the constant external
fields are eigenstates of the Hamiltonian $ \hat H _ 0 $ in the
absence of wave: $ \hat H _ 0\mid n> = \varepsilon _ n \mid n> $.
Expansion of the state $ \mid t> $ to states  $ \mid n> $ gives
\begin {equation}
\mid t> = \mid 0> + \sum _ {n\neq 0} a _ n (t) | n> e ^
{-i\varepsilon _ n t}.
\end {equation}
Substituting the given expression to Eq. (\ref {sh1}) we obtain:
\begin {equation}
i\frac {da _ n (t)} {dt} = < n | \hat V (t) | 0> e ^ {i\varepsilon
_ n t}. \label {eva}
\end {equation}
Using the Fourier transform of the wave 4-potential $ A _\mu(x)=
\int A _ \mu (k) e ^ {-ikx} d ^ 4x $ and the translational
invariance of vacuum $ < n | \hat j _ \mu (\bm r) | 0> = < n |
\hat j _ \mu (0) | 0> e ^ {-i\bm p _ n\bm r} $ we obtain
\begin{widetext}
\begin {eqnarray}
< n | \hat V (t) | 0> = \int < n | \hat j ^ \mu (\bm r) | 0> e ^
{i\varepsilon _ nt-ikx} A _ \mu (k) d ^ 4kd ^ 3\bm r
= \int < n |\hat j ^ \mu (0) | 0> e ^ {ip _ nx-ikx} A _ \mu (k) d ^ 4kd ^ 3\bm r\nonumber \\
= (2\pi) ^ 3 < n | \hat j ^ \mu (0) | 0> \int\delta ^ {(3)} (\bm p
_ n-\bm k) e^ {i\varepsilon _ nt-i\omega t} A _ \mu (k) d ^ 4k.
\end {eqnarray}
\end{widetext}
The solution of  Eq. (\ref {eva}) can be written as
\begin{widetext}
\begin {equation}
a_ n (t) = -i\int ^ t _ {-\infty} < n | \hat V (\tau) | 0> e^
{i\varepsilon _ n \tau} d\tau
 = (2\pi)^3 < n | \hat j^\mu (0) | 0> \int\delta^{(3)} (\bm p _ n + \bm k)
 \frac {e ^ {i\varepsilon _ nt-i\omega
 t}} {\omega-\varepsilon _ n + i0} A _ \mu (k) d ^ 4k.
\label{an}
\end {equation}
\end{widetext}
Then we can find the average value of $\hat j(\bm r)$. Evaluation
of $j(\bm r, t)=<t\mid \hat j(\bm r)\mid t>$ with the help of
$a_n(t)$ given by Eq. (\ref{an}) leads to
\begin{widetext}
\begin{eqnarray}
j_\mu(x)=-(2\pi)^3\int\sum_{n\neq 0}\biggl(<n|\hat
j^\nu(0)|0>\delta^{(3)}(\bm p_n-\bm
k)\frac{e^{i(\varepsilon_n-\omega)t}}{\varepsilon_n-\omega-i0}A_\nu(k)
<0|\hat j_\mu(x)|n> \nonumber \\ + <0|\hat
j^\nu(0)|n>\delta^{(3)}(\bm p_n-\bm
k)\frac{e^{-i(\varepsilon_n-\omega)t}}{\varepsilon_n-\omega+i0}A^*_\nu(k)
<n|\hat j_\mu(x)|0> \biggr)d^4k \nonumber
\\
=-(2\pi)^3\int\sum_n \biggl(<0|\hat j_\mu(0)|n><n|\hat
j^\nu(0)|0>\frac{\delta^{(3)}(\bm p_n-\bm
k)}{\varepsilon_n-\omega-i0} \nonumber \\ +<0|\hat
j^\nu(0)|n><n|\hat j_\mu(0)|0>\frac{\delta^{(3)}(\bm p_n+\bm
k)}{\varepsilon_n+\omega+i0}\biggr)A_\nu(k)e^{-ikx}d^4k.
\label{tok1}
\end{eqnarray}
\end{widetext}
From Eq. (\ref{tok1}), in view of the definition given by Eq.
(\ref {2}) we obtain Eq. (\ref {112}), which is the Fourier
transform
 of Eq. (\ref {555}). Let's note, that the Fourier transform
of the causal polarization operator
\begin{equation}
\Pi_{\mu\nu}(x)=4\pi i <0|T\hat j_\mu(x)\hat j_\nu(0)|0>
\end{equation}
differs from (\ref {112}) by the sign before $i0$ in the second
term. For the photon refractive index it is necessary to use just
the delayed  polarization operator $ {\mathcal P}_{\mu\nu} $, as
in this case ${\mathcal P }^{\mu\nu}(k)$ given by Eq. (\ref {112})
has the right properties $ {\mathcal P}^*_{\mu\nu} (k) = {\mathcal
P}_{\mu\nu} (-k) $ required by a reality of the field $A(x)$.

\end{document}